\documentclass[preprint,12pt]{elsarticle}

\usepackage{graphics}

\usepackage{amssymb}
\usepackage{latexsym}

\journal{Physica A}

\begin{document}

\begin{frontmatter}



\title{
Magnon-bound-state
hierarchy for 
the two-dimensional transverse-field Ising model
in the ordered phase
}


\author{Yoshihiro Nishiyama} 

\address{Department of Physics, Faculty of Science,
Okayama University, Okayama 700-8530, Japan}

\begin{abstract}
In the ordered phase for an Ising ferromagnet,
the magnons are attractive
to
form a series of bound states with
the
mass gaps,
$m_2<m_3 < \dots$.
Each ratio $m_{2,3,\dots}/m_1$ 
($m_1$: the single-magnon mass)
is expected to be a universal constant
in the vicinity of the critical point.
In this paper,
we devote ourselves to the
$(2+1)$-dimensional counterpart,
for which 
the universal hierarchical character remains unclear.
We employed the exact diagonalization method,
which enables us to calculate the dynamical susceptibility
via the continued-fraction expansion.
Thereby, we observe 
a variety of signals including $m_{2,3,4}$,
and the spectrum is analyzed with the finite-size-scaling method
to estimate the universal mass-gap ratios.

\end{abstract}


\end{frontmatter}



\section{\label{section1}Introduction}

For an Ising ferromagnet in the ordered phase,
the magnons are attractive,
forming 
a series of bound states with the mass gaps,
$m_2<m_3 <\dots$.
As a matter of fact,
in $(1+1)$ dimensions, where 
the system is integrable
\cite{Zamolodchikov98},
there exist
eight types of elementary excitations with
the universal mass gaps
\begin{eqnarray}
m_2/m_1 &=& 2 \cos {\pi}/{5}\nonumber \\
m_3/m_1 &=& 2\cos {\pi}/{30}\nonumber \\
m_4/m_1 &=& 4 \cos {\pi}/{5}\cos{7\pi}/{30}\nonumber \\
m_5/m_1 &=& 4\cos {\pi}/{5}\cos{2\pi}/{15}\nonumber \\
m_6/m_1 &=& 4\cos {\pi}/{5}\cos{\pi}/{30}\nonumber \\
m_7/m_1 &=& 8\cos^2 {\pi}/{5}\cos{7\pi}/{30}\nonumber \\
m_8/m_1 &=& 8 \cos^2{\pi}/{5}\cos{2\pi}/{15} ,  \nonumber 
\end{eqnarray}
($m_1$: single-magnon mass gap)
in the vicinity of the critical point
\cite{Delfino04,Fonseca03}. 
According to the rigorous theory,
the {\em elementary} magnon $m_1$ is also a composite particle,
reflecting a highly non-perturbative character
of this problem;
in this sense,
the underlying physics
may lie out of the 
conventional ``magnon" picture.
Experimentally,
the lowest one $m_2/m_1=1.618 \dots$
(golden ratio)
was observed
\cite{Coldea10}
for a quasi-one-dimensional quantum Ising ferromagnet, CoNb$_2$O$_6$
\cite{Lee10,Ishimura80}.
Above $\omega / m_1 \ge 2$,
there extends a two-magnon continuum,
overwhelming fine details of the spectrum;
see Fig. 4 E of Ref. \cite{Coldea10}, for instance.

\begin{figure}
\includegraphics[width=100mm]{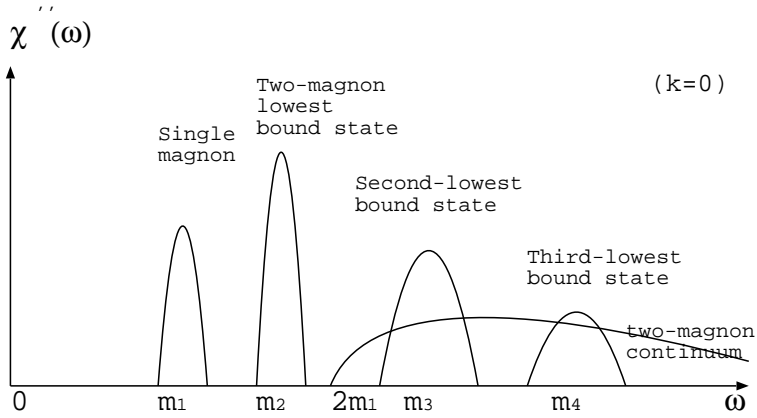}%
\caption{
\label{figure1}
A schematic drawing of the
spectral function
for the $(2+1)$-dimensional Ising model
in the ordered phase
at the zero-momentum $k=0$ sector
is shown.
The spectrum 
is expected to exhibit a universal character
in the vicinity of the critical point.
The $m_1$ peak corresponds to the single-magnon excitation.
The peaks at $m_{2,3,\dots}$ are the bound states,
which may be embedded within the two-magnon continuum
extending
above $2 m_1$.
}
\end{figure}

In $(2+1)$ dimensions,
on the contrary,
such rigorous information is not available,
and details 
of the bound-state hierarchy are
not fully clarified;
the role of dimensionality 
was argued in \S 5 of Ref. \cite{Caselle02} 
(see Ref. \cite{Rose16} as well).
To the best of author's knowledge,
the second-lowest bound state $m_3/m_1 = 2.45(10)$ 
was detected with the Monte Carlo method
\cite{Caselle99}, whereas 
the lowest one, 
$m_2/m_1 \approx 1.81$ \cite{Dusuel10},
has been investigated rather extensively so far
\cite{Caselle99,Lee01,Caselle02,Nishiyama08,Nishiyama14,Rose16}.
In this paper, 
we investigate the $(2+1)$-dimensional
Ising model (\ref{Hamiltonian})
with the exact diagonalization method,
which enables us to calculate the dynamical susceptibility 
via the continued-fraction expansion \cite{Gagliano87};
note that in the Monte Carlo simulation,
one has to
resort to the inverse Laplace transform to obtain the spectrum.
The spectrum 
reflects a hierarchical character for $m_{2,3,\dots}$.
In Fig. \ref{figure1},
we present a schematic drawing  for a spectral function
within the zero-momentum sector.

It has to be mentioned that the 
magnon-bound-state hierarchy 
is relevant to the
glueball spectrum
(screening masses)
for the gauge field theory
(Svetitsky-Yaffe conjecture)
\cite{Svetitsky82,Agostini97,Fiore03}.
Actually, 
we show that in the next section,
the bound-state hierarchy $m_{2,3,4}$
bears a resemblance to
the glueball spectrum
for the Z$_2$-symmetric gauge field theory \cite{Agostini97}.
Here, we dwell on the characterization of the magnon bound states,
and the verification of the conjecture itself
lies beyond the scope
of this paper.

\begin{figure}
\includegraphics[width=100mm]{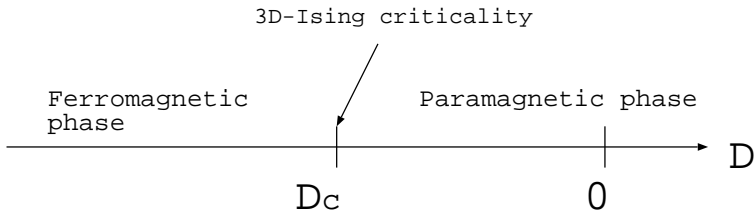}%
\caption{
\label{figure2}
The phase diagram for the
two-dimensional 
transverse-field Ising model 
(\ref{Hamiltonian}) is presented;
here, the single-ion anisotropy $D$ is a variable parameter,
and the other interaction parameters are set to 
Eq. (\ref{fixed_point})
so as to suppress corrections to scaling \cite{Nishiyama10}.
As the parameter $D$ varies,
a phase transition separating the ferromagnetic and paramagnetic
phases takes place at $D=D_c$. 
}
\end{figure}

To be specific,
we
present the Hamiltonian
for the two-dimensional 
spin-$S=1$ transverse-field Ising
model
\begin{eqnarray}
\label{Hamiltonian}
{\cal H}         & = & 
-J \sum_{\langle ij \rangle} S^z_i S^z_j
- J'\sum_{\langle \langle ij \rangle \rangle} S^z_i S^z_j
- J_4 \sum_{\langle ij \rangle} (S^z_i S^z_j)^2  \nonumber \\
 & &
- J_4' \sum_{\langle \langle ij \rangle \rangle} (S^z_i S^z_j)^2
  +D \sum_{i} (S^z_i)^2
     -\Gamma \sum_i S^x_i
   -H  \sum_i S^z_i
              , 
\end{eqnarray}
with 
the
quantum spin-$S=1$ operator ${\bf S}_i$ placed at each square-lattice point $i$.
The summations,
 $\sum_{\langle ij \rangle}$
and 
 $\sum_{\langle \langle ij \rangle \rangle}$,
run
over all possible nearest-neighbor 
and next-nearest-neighbor
pairs, 
$\langle ij \rangle$ and 
$\langle \langle ij \rangle \rangle$,
respectively.
Correspondingly,
$J$ ($J_4$)
and
$J'$ ($J'_4$)
  are the quadratic (biquadratic)
interaction parameters.
The symbols,
$D$,
$\Gamma$ and $H$,
denote
the single-ion anisotropy,
the transverse- and longitudinal-magnetic fields,
respectively.
The parameter $D$ is tunable,
and the phase diagram 
is presented in Fig. \ref{figure2}.
Other 
interaction parameters are
set to
\begin{eqnarray}
(J,J',J_4,J_4',\Gamma) 
&=& [
 0.4119169708 5 ,  
0.1612506961 6 ,  
-0.1176402001 8 ,    \nonumber \\  
\label{fixed_point}
& & 
-0.0 526792660 1 ,  
1.0007    
]
%
,
\end{eqnarray}
so as to suppress corrections to scaling \cite{Nishiyama10}.
The 
(properly scaled)
infinitesimal magnetic field $H=11 L^{-y_h}$ \cite{Nishiyama10}
with
$y_h=2.481865$ \cite{Hasenbusch10}  
resolves the ground-state
two-fold degeneracy \cite{Fonseca03}.
The $S=1$-spin model permits us to incorporate extended
interactions so as to suppress corrections to scaling
\cite{Deng03}.
In fact, as demonstrated in Ref. \cite{Nishiyama10},
even for restricted system sizes,
the Hamiltonian
(\ref{Hamiltonian}) exhibits 
suppressed corrections to scaling.
Detailed account of
the three-dimensional Ising universality
is reported
in Ref. 
\cite{Hasenbusch10}.

The rest of this paper is organized as follows.
In Sec. \ref{section2}, 
we present the numerical results.
The simulation algorithm is explained as well.
In Sec. \ref{section3}, we
address the summary and discussions.

\section{\label{section2}Numerical results}

In this section,
we present the numerical results
for the $(2+1)$-dimensional Ising model (\ref{Hamiltonian}).
We employed the exact diagonalization method for 
the finite-size cluster with $N \le 22$ spins.
We imposed the screw-boundary condition 
\cite{Novotny90}
to treat 
an arbitrary number of spins $N=16,18,\dots$.
We adopt
the simulation algorithm presented in
Appendix of 
Ref. \cite{Nishiyama10}.
Because the $N$ spins constitute a rectangular cluster,
the linear dimension of the cluster is given by the formula
\begin{equation}
L=\sqrt{N} .
\end{equation}
The diagonalization was performed 
within the zero-momentum subspace.

\subsection{\label{section2_1}
Critical behavior of the
single-magnon mass gap
$m_1$}

In this section,
we make a finite-size-scaling analysis for the
single-magnon mass gap $m_1$,
which sets a fundamental energy scale
in the subsequent scaling analyses.
The single-magnon mass $m_1$
corresponds to the first excitation gap
above the ground state.
(The ground-state two-fold degeneracy is resolved by $H$,
as mentioned in Sec. \ref{section1}.)

\begin{figure}
\includegraphics[width=100mm]{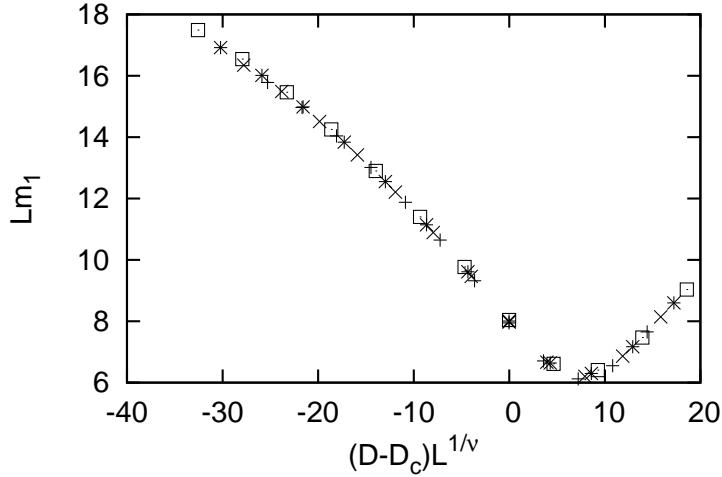}%
\caption{
\label{figure3}
The scaling plot,
$(D-D_c)L^{1/\nu}$-$Lm_1$,
is presented for 
($+$) $N =16$,
($\times$) $18$,
($*$) $20$,
and ($\Box$) $22$.
Here, the scaling parameters, 
$D_c=-0.3978195612 2 $ 
 and 
$\nu=0.63002$, 
are taken from Refs.
 \cite{Nishiyama10}
 and 
 \cite{Hasenbusch10}, respectively;
namely, there is no adjustable fitting parameter
involved
in the scaling analysis.
}
\end{figure}

In Fig. \ref{figure3},
we present the scaling plot, $(D-D_c)L^{1/\nu}$-$L m_1$, for the
number of spins, 
$(+)$ $N=16$,
$(\times)$ $18$,
$(*)$ $20$, and 
$(\Box)$ $22$.
Here, the scaling parameters, 
$D_c=-0.3978195612 2 $ 
and 
$\nu=0.63002$, 
are taken from the existing literatures,
Refs.
\cite{Nishiyama10} and 
\cite{Hasenbusch10},
respectively.
That is,
there is no adjustable fitting parameter
in the scaling analysis.
We observe that 
the data in Fig. \ref{figure3}
collapse into a scaling curve for
a considerably wide range of the parameter $D$.
The simulation data already enter the scaling regime.
Encouraged by this finding,
we turn to the analysis of the spectral properties.

Last, we address a few remarks.
First,
the data in Fig. \ref{figure3}
indicate that the gap closes beside the critical point,
and the gap itself seems to be rather large.
These confusing features are due to the infinitesimal (scaled)
longitudinal magnetic field as mentioned in Introduction.
This infinitesimal magnetic field resolves the ground-state degeneracy,
stabilizing the magnetic excitations \cite{Nishiyama10}.
Last,
right at the critical point,
the excitation hierarchy becomes smeared out.
The smearing-out regime
extends
within a bound
 $|D-D_c| \propto 1/L^{1/\nu}$,
 which shrinks in the thermodynamic limit.

\subsection{\label{section2_2}
Finite-size-scaling analysis of the
dynamical susceptibility
$\chi_{Y^2}''(\omega)$:
Magnon-bound-state hierarchy}

\begin{figure}
\includegraphics[width=100mm]{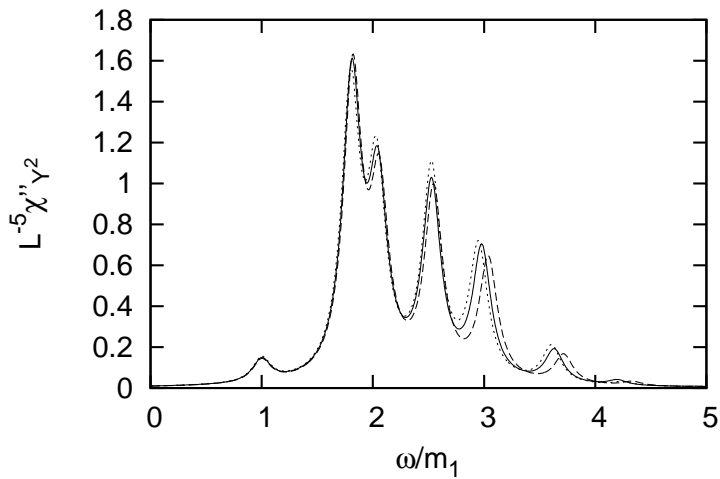}%
\caption{
\label{figure4}
The scaling plot,
$\omega/m_1$-$L^{-5} \chi_{Y^2}''(\omega)$,
is shown with $(D-D_c)L^{1/\nu}=-17$
and
$\eta=0.1m_1$
for $N=18$ (dotted),
$20$ (solid),
and
$22$ (dashed).
The scaling parameters, 
$D_c$
and $\nu$, are
the same as those of Fig. \ref{figure3}.
The character of each spectral peak
is argued in the text.
}
\end{figure}

In this section,
we investigate
the magnon-bound-state hierarchy.
For that purpose,
we introduce 
the dynamical susceptibility
\begin{equation}
\label{susceptibility1}
\chi_{Y^2}'' (\omega)
= - \Im  
      \langle 0 |M_{Y^2}^\dagger
(\omega-{\cal H}+E_0+{\rm i}\eta)^{-1}
M_{Y^2}
 |0 \rangle  ,
\end{equation}
with the perturbation operator
\begin{equation}
\label{perturbation1}
M_{Y^2}={\cal P} \left(
     \sum_{i=1}^{N} S^y_i
           \right)^2
  ,
\end{equation}
and the projection operator 
${\cal P}=1-| 0 \rangle\langle 0 |$.
Here, the symbols,
$\eta$ and
$E_0$ 
($|0\rangle$), 
denote the energy-resolution parameter
and the
ground-state energy (vector),
respectively.
The dynamical susceptibility 
was calculated with the continued-fraction expansion
\cite{Gagliano87}.
Rather technically, the continued-fraction expansion 
requires the iteration sequences less than those
of the Lanczos diagonalization
for $E_0$ and $|0\rangle$; namely, at least for drawing the spectra such as 
Fig. \ref{figure4}-\ref{figure7},
the results converge rapidly
possibly owing to the plausible choice of $M_{Y^2}$.
The dynamical susceptibility obeys
\cite{Podolsky11}
the scaling formula
\begin{equation}
\label{scaling_formula}
\chi_{Y^2}'' = L^5 g(\omega/m_1 , (D-D_c)L^{1/\nu} )
	                ,
\end{equation}
with a scaling function $g$.
Afterward,
the power-law singularity $\sim L^5$
as well as the physical content of $\chi_{Y^2}$
are considered.

In Fig. \ref{figure4},
we present the scaling plot, 
$\omega / m_1$-$L^{-5} \chi_{Y^2}''$,
with fixed
$(D-D_c) L^{1/\nu}=-17$ and $\eta=0.1m_1$
for various
$N=18 $ (dotted),
$20$ (solid),
and $22$ (dashed).
The scaling parameters, $D_c$ and $\nu$,
are the same as those of Fig. \ref{figure3}. 
The curves in Fig. \ref{figure4}
collapse into a scaling function $g$, Eq.
(\ref{scaling_formula}),
indicating that the simulation data already enter the scaling regime.

The peaks in Fig. \ref{figure4}
reflect a universal character of the magnon excitations
in proximity to the critical point;
see a schematic drawing, Fig. \ref{figure1}, as well.
The peaks at 
$\omega/m_1 \approx 1$,
$1.8$,
$2$,
$2.5$
and
$3$
correspond to the
$m_1$,
$m_2$ \cite{Caselle99,Dusuel10,Caselle02,Nishiyama08,Nishiyama14,Rose16},
$2m_1$
(two-magnon-continuum threshold),
$m_3$ \cite{Caselle99},
and
$m_4$
excitations, respectively.
The $\omega/m_1  \approx 4.8$-peak
may be attributed to either 
a yet higher bound state
 $m_5$ 
or
a 
composite particle made of $m_2$
and $2 m_1$.

It has to be mentioned that
these excitations $m_{2,3,4}$
correspond to the glueball spectrum
(screening masses) for the gauge field theory
(Svetitsky-Yaffe conjecture \cite{Svetitsky82}).
For instance,
according to the Monte Carlo simulation of
the 
${\rm Z}_2$ lattice gauge-field theory
\cite{Agostini97},
there appear a variety of particles (the so-called glueballs),
$m_{(0^+)'}/m_{0^+}=1.88(2)$,
$m_{(0^+)''}/m_{0^+}=2.59(4)$,
and
$m_{(0^-)}/m_{0^+}=3.25(16)$,
quite reminiscent of the above-mentioned $m_{2,3,4}/m_1$, respectively;
here we follow the notation of Ref. 
\cite{Agostini97}.
Our simulation result supports
the Svetitsky-Yaffe conjecture 
at least up to $m_4$.

A few remarks are in order.
First, we consider the power-law singularity of 
$\chi_{Y^2} \sim L^5 $, Eq.(\ref{scaling_formula}).
By analogy to the random-walk translation distance,
the singularity of
the perturbation
operator $M_{Y^2}=(\sum_{i=1}^N S^y_i)^2$,
Eq.
(\ref{perturbation1}), is counted as
$ \sim N(=L^2)$.
On the one hand,
the reciprocal energy gap should exhibit the singularity,
$(\omega-{\cal H}+E_0+{\rm i}\eta)^{-1} \sim  L$.
These formulas lead to 
$\chi_{Y^2}'' \sim L^5$.
As a matter of fact, there also appears
a non-trivial term $\chi_{Y^2}\sim L^{2/\nu-3}$;
this term
is less singular than the above $\sim L^5$,
and it was dropped in Eq. (\ref{scaling_formula}).
Here,
a key ingredient is that an infinitesimal 
$y$-direction magnetic field $H^y$
is perpendicular to $\Gamma$,
and hence,
the scaling dimension of $H^y$
is a half of $\Gamma$.
Second,
we make an overview of the dynamical susceptibilities
appearing in literature.
So far, there have been utilized two types of perturbations,
namely, off-diagonal
$M_{Y}=\sum_{i=1}^N S^y_i$ and 
diagonal
$M_{Z}=\sum_{i=1}^N S^z_i$ ones,
in the course of study for both
$(1+1)$- and $(2+1)$-dimensional systems.
The former
$M_Y$ was employed in Refs. 
\cite{Seabra13,Kjall11,Robinson14,Wang15}.
Our choice $M_{Y^2}$ (\ref{perturbation1}) is 
based on
these elaborated studies.
We found that
the duplicated operations $(M_{Y})^2=M_{Y^2}$ create 
the $m_{3,4}$ particles 
more efficiently 
than a mere $M_Y$, at least, for the $(2+1)$-dimensional counterpart.
The latter
$M_Z$
was implemented in Refs. 
\cite{Robinson14,Wang15,Delfino95}.
This choice leads to the ordinary uniform magnetic AC susceptibility,
and it is of experimental significance.
Last,
we mention the parameter setting of
the
peak-broadening factor $\eta$.
In this paper, aiming to avoid 
{\it ad hoc} adjustment of $\eta$,
we fixed the value of $\eta$ throughout the study.
The continuum spectrum in Fig. \ref{figure7} seems to be
well reproduced by this setting.

\subsection{\label{section2_3}
Universality of the bound-state hierarchy
 $m_{2,3,4}/m_1$}

In the above section,
through the probe $\chi_{Y^2}''$,
we resolved the $m_{2,3,4}$ signals out of the two-magnon continuum.
In this section, we examine the universality
 of each mass-gap ratio
$m_{2,3,4}/m_1$ with respect to the variation of
$(D-D_c)L^{1/\nu}$.

\begin{figure}
\includegraphics[width=100mm]{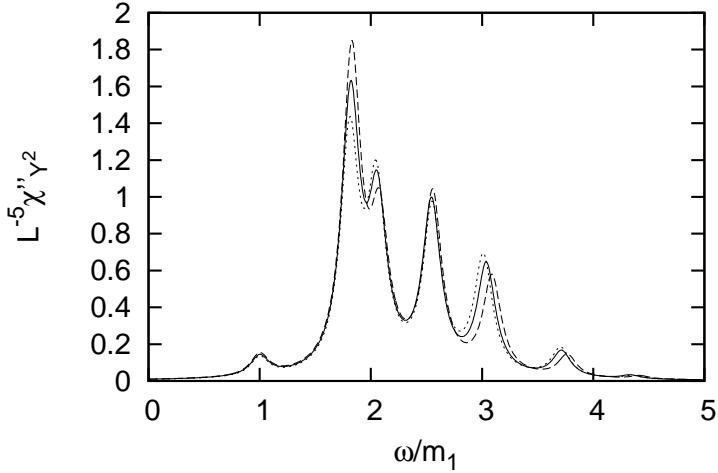}%
\caption{
\label{figure5}
The scaling plot,
$\omega/m_1$-$L^{-5}\chi_{Y^2}''(\omega)$,
is shown with $N=22$ and
$\eta=0.1m_1$
for various values of the scaling argument,
$(D-D_c)L^{1/\nu}=-15$ (dashed),
$-17$ (solid),
and $-19$ (dotted);
the scaling parameters,
$D_c$ and $\nu$, are the same as those of Fig. \ref{figure3}.
These curves by no means overlap,
because the scaling argument is ranging;
the heights of the $m_2$ peak are scattered,
whereas
the position $m_2/m_1\approx 1.8 $ 
is kept invariant.
}
\end{figure}

\begin{figure}
\includegraphics[width=100mm]{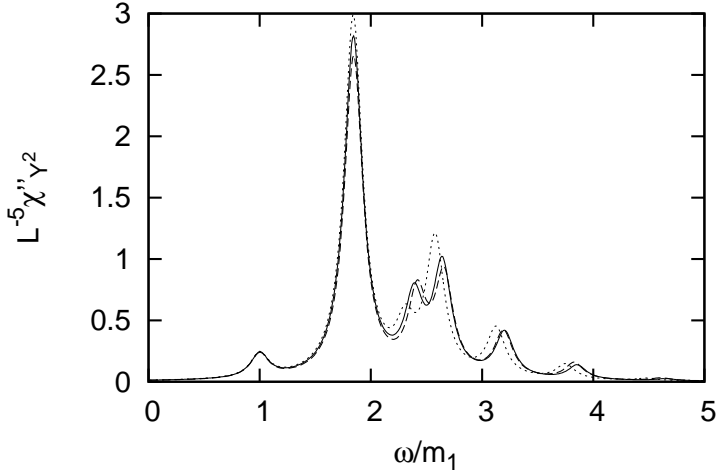}%
\caption{
\label{figure6}
The scaling plot,
$\omega/m_1$-$L^{-5}\chi_{Y^2}''(\omega)$,
is shown with $(D-D_c)L^{1/\nu}=-5.5$ and
$\eta=0.1m_1$
for
$N=18$ (dotted),
$20$ (solid), and 
$22$ (dashed);
the scaling parameters,
$D_c$ and $\nu$, are the same as those of Fig. \ref{figure3}.
In this regime,
the $m_3$ peak splits
into the sub-peaks at $\omega / m_1 \approx 2.41$ and $2.65$ ($N=22$),
indicating that the $m_3$ peak is broadened 
intrinsically.
}
\end{figure}

In Fig. \ref{figure5},
we present the scaling plot,
 $\omega/m_1$-$L^{-5}\chi_{Y^2}''$,
with $N=22$ and $\eta=0.1m_1$
for various 
values of the scaling argument,
$(D-D_c)L^{1/\nu}=-15$ (dashed), 
$-17$ (solid),
and 
$-19$ (dotted);
here, the scaling parameters, $D_c$ and $\nu$,
are
 the same as those of Fig. \ref{figure3}.
Note that the curves do not necessarily overlap,
because the
scaling argument
$(D-D_c)L^{1/ \nu}$
is ranging; see Eq. (\ref{scaling_formula}).
In fact, 
the $m_2$-peak heights do not coincide each other,
whereas
the peak position,
 $m_2/m_1 \approx 1.81$
\cite{Dusuel10}, appears 
to be kept invariant with 
$(D-D_c)L^{1/ \nu}$ varied.
Our result 
supports the universality of $m_2/m_1 \approx 1.81$ \cite{Dusuel10},
which has been
established in the course of study
\cite{Caselle99,Caselle02,Nishiyama08,Nishiyama14,Rose16}.

We turn to the analysis of the second-lowest bound state, $m_3/m_1$.
In Fig. \ref{figure5}, we observe that the peak position $m_3/m_1$ 
is kept invariant with $(D-D_c)L^{1/\nu}$ varied.
In closer look ($N=22$),
the peak positions read
$m_3/m_1 \approx 2.56$,
$2.55$,
and $2.56$ for 
$(D-D_c)L^{1/\nu}=-15$,
$-17$, 
and $-19$,
respectively.
Namely,
the condition 
$(D-D_c)L^{1/\nu}=-17$ is an
optimal one
in the sense that the peak position $m_3/m_1$ takes 
a stable (extremal)
value
\begin{equation}
\label{m3}
m_3/m_1=2.55 .
\end{equation}
Our result agrees with the Monte Carlo result
$m_3/m_1=2.45(10)$ 
\cite{Caselle99}.

Additionally,
we are able to appreciate the
universal intrinsic
peak width for $m_3$. 
The $m_3$ particle may
decay into a pair of $m_1$,
and eventually,
the peak
acquires 
a finite life time,
namely, a reciprocal intrinsic peak width;
because the $(2+1)$-dimensional system is not
integrable, it is natural that the spectral intensities
are diffused
\cite{Robinson14}.
In Fig. \ref{figure6},
we present the scaling plot, 
$\omega/m_1$-$L^{-5}\chi_{Y^2}''$,
with $(D-D_c)L^{1/\nu}=-5.5$ and $\eta=0.1m_1$
for 
$N=18$ (dotted),
$20$ (solid), and
$22$ (dashed);
the 
scaling parameters, $D_c$ and $\nu$, are the same as those of Fig. \ref{figure3}. 
In this scaling regime,
 $(D-D_c)L^{1/\nu}=-5.5$,
the $m_3$ peak splits 
into two sub-peaks at $\omega/m_1\approx 2.41$ and $2.65$
($N=22$).
The distance between these sub-peaks 
provides an indicator for
the intrinsic peak width 
\begin{equation}
\label{delta_m3}
\delta m_3/m_1 =0.24  .
\end{equation}
The peak width is smaller by one order of magnitude than
the mass (\ref{m3}),
indicating that the $m_3$ particle is a stable collective mode,
and it would be observable experimentally.
We consider that the error margin for $m_3/m_1=2.55$,
Eq.
(\ref{m3}), would not exceed a half of this peak width;
namely, the uncertainty for $m_3/m_1=2.55$,
Eq. (\ref{m3}),
is bounded by 
 $0.12$.

Last,
we consider the $m_4$ signal.
From Fig. \ref{figure5},
we estimate the mass-gap ratio as $m_4/m_1 \approx 3$.
The peak position appears to drift,
indicating that this peak consists of diffused intensities.
In fact, there are a number of decay modes
such as $m_4 \to 2m_1$, $m_1+m_2$ and $3m_1$,
giving rise to a considerably broadened line shape 
as to $m_4$.
Further details are not fixed
  by the available result.

We address a remark on the $D$-dependence of the spectrum.
As mentioned above, the peak position takes
a stable (extremal) value at an
optimal condition $(D-D_c)L^{1/\nu}=-17$.
In a closer look, however, there are irregularities
caused by the level crossing.
A level-crossing point locates
at the above-mentioned point $(D-D_c)L^{1/\nu}=-5.5$,
and another one occurs around $(D-D_c)L^{1/\nu}=-14$, where the data of $N=20$ and $22$
are not very influenced.
In this sense,
these intermittent irregularities are due to the finite-size artifact,
reflecting the fact that the bound state is embedded within the continuum.
As mentioned above, the level crossing provides information
how the spectral intensity is distributed over the discrete levels,
and hence, it indicates the intrinsic peak width of 
the spectral peak concerned.
Nevertheless, apart from these intermittent irregularities,
the peak position is kept invariant for a considerably wide range of $D$,
and the slight wavy deviation appears to be bounded by 
the above-mentioned error margin. 

\subsection{\label{section2_4}
Finite-size-scaling analysis of the
dynamical susceptibility
$\chi_{z^2}''(\omega)$:
Two-magnon continuum}

In this section,
we investigate the dynamical susceptibility
$\chi_{z^2}''$, Eq. (\ref{susceptibility2});
here,
the perturbation operator $M_{z^2}$ (\ref{perturbation2})
is identical to the $D$ term of the Hamiltonian (\ref{Hamiltonian}),
and such a dynamical susceptibility
is called the scalar susceptibility \cite{Podolsky11}
in literature.

\begin{figure}
\includegraphics[width=100mm]{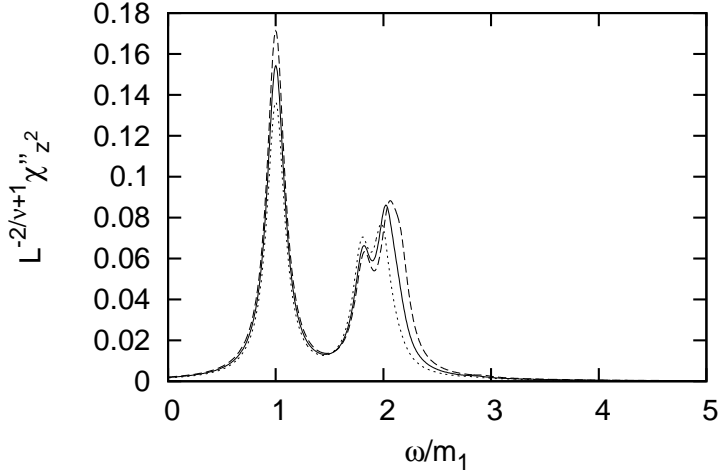}%
\caption{
\label{figure7}
The scaling plot,
$\omega/m_1$-$L^{-2/\nu+1}\chi_{z^2}''(\omega)$,
is shown with $(D-D_c)L^{1/\nu}=-17$ and
$\eta=0.1m_1$
for
$N=18$ (dotted),
$20$ (solid),
and $22$ (dashed);
the scaling parameters,
$D_c$ and $\nu$, are the same as those of Fig. \ref{figure3}.
The susceptibility
$\chi_{z^2}''$
is sensitive
to the two-magnon continuum 
extending above $\omega/m_1 \ge 2$,
which overwhelms
the $m_{3,4}$ peaks.
}
\end{figure}

In Fig. \ref{figure7},
we present the scaling plot,
$\omega/m_1$-$L^{-2/\nu+1}\chi_{z^2}''$,
with $(D-D_c)L^{1/\nu}=-17$ 
and
$\eta=0.1m_1$
for $N=18$ (dotted),
$20$ (solid),
and $22$ (dashed);
the scaling parameters, 
$D_c$
and $\nu$,
are the same as those of Fig. \ref{figure3}.
The dynamical susceptibility 
is defined by the formula
\begin{equation}
\label{susceptibility2}
\chi_{z^2}'' (\omega) = - \Im \langle 0|
M_{z^2}^\dagger (\omega-{\cal H} +i\eta)^{-1} 
M_{z^2}
|0\rangle   ,
\end{equation}
with the perturbation operator
\begin{equation}
	\label{perturbation2}
	M_{z^2}= {\cal P}\sum_{i=1}^N (S^z_i)^2
	      .
\end{equation}
The dynamical susceptibility
obeys the scaling law
$\chi_{z^2}''=L^{2/\nu-1}h(\omega/m_1 , (D-D_c)L^{1/\nu} )$ \cite{Podolsky11}
with  a scaling function $h$.

The data
in Fig. \ref{figure7},
indicate that this probe
$\chi_{z^2}''$ is sensitive to the two-magnon
continuum, particularly,
its threshold $\omega/m_1 \approx 2$.
Actually, in contrast to Fig. \ref{figure4},
the bound-state hierarchy $m_{3,4}$ becomes invisible instead.
Our simulation result suggests that
the
choice of the perturbation operator
is vital to the detection of
the bound-state hierarchy.

We address a remark.
The
scatter of the scaling curves in Fig. \ref{figure7}
should be attributed to the regular part (non-singular contribution)
of $\chi_{z^2}''$.
Because the leading singularity of $\chi_{z^2}''\sim L^{2/\nu-1}$
is rather weak, it is contaminated by such residual scaling corrections.

\section{\label{section3}
Summary and discussions}

The universal character of the
magnon-bound-state hierarchy for the two-dimensional
quantum Ising model (\ref{Hamiltonian}) in the ordered phase
was investigated.
So far, the lowest bound state $m_2$ has been investigated 
rather extensively
\cite{Caselle99,Dusuel10,Lee01,Caselle02,Nishiyama08,Nishiyama14,Rose16}.
We employed the exact diagonalization method,
which yields the dynamical susceptibilities
$\chi_{Y^2,z^2}''$ via the continued-fraction expansion 
\cite{Gagliano87}.
The dynamical susceptibility $\chi_{Y^2}''$
resolves the bound-state hierarchy $m_{2 , 3, 4}$
out of the two-magnon continuum.
Making the finite-size-scaling analysis of $\chi_{Y^2}''$,
we estimated the mass-gap ratios.
The lowest one agrees with the
preceding result
$m_2/m_1=1.81$ \cite{Dusuel10},
which has been established in the course of study
\cite{Caselle99,Caselle02,Nishiyama08,Nishiyama14,Rose16}.
For the second-lowest bound state,
we estimated the mass-gap ratio,
$m_3/m_1=2.55$.
Our result agrees with the 
Monte Carlo result $m_3/m_1=2.45(10)$
\cite{Caselle99}.
Additionally, we appreciated its
intrinsic peak width, $\delta m_3/m_1=0.24$;
 the bound state $m_3$ exhibits a rather long life time
as a stable collective mode,
though it is embedded within the continuum.
A signature for the third-lowest 
bound state is also detected around $m_4 / m_1 \approx 3$.
In contrast to $\chi_{Y^2}''$,
the probe $\chi_{z^2}''$ 
is sensitive to the two-magnon continuum,
which smears out the $m_{3,4}$ signals eventually.
The choice of the perturbation field may be significant
for resolving the magnon bound states out of the
continuum.

As demonstrated in Sec. \ref{section2_3},
the bound state hierarchy $m_{2,3,4}$
bears a resemblance to
the glueball spectrum
$m_{(0^+)',(0^+)'',(0^-)}$ \cite{Agostini97},
supporting the Svetitsky-Yaffe conjecture 
\cite{Svetitsky82}.
It is tempting to apply the similar approach
to the $q$-state Potts model 
in order to examine the validity of the conjecture.
This problem is left for the future study.

\section*{Acknowledgment}
This work was supported by a Grant-in-Aid 
for Scientific Research (C) from Japan Society for
the Promotion of Science
(Grant No. 25400402).

\bibliographystyle{elsarticle-num}







\end{document}